# Significance of Disk Failure Prediction in Datacenters

JAYANTA BASAK, NetApp Inc.
RANDY H. KATZ, University of California, Berkeley

Modern datacenters assemble a very large number of disk drives under a single roof. Even if economic and technical factors where to make individual drives more reliable (which is not at all clear, given the commoditization of the technology), their sheer numbers combined with their ever increasing utilization in a well-balanced design makes achieving storage reliability a major challenge. In this paper, we assess the challenge of storage system reliability in the modern datacenter, and demonstrate how good disk failure prediction models can significantly improve the reliability of such systems.

## 1. INTRODUCTION

"Big data" imposes new challenges for datacenter storage. Datacenters are still mostly dominated by hard disk storage. Recent studies reveal that SSDs are also vulnerable to failures [Schroeder et al. 2016]. Google wants to see storage technology evolve to meet the demands of cloud computing by adopting less reliable hard disks [Brewer et al. 2016] and Google envisages new hard disk format design for data centers. As storage capacity and the number of disks increase, data loss becomes not just a possibility but inevitable. The sheer size and scale of data movement between applications and storage in a large-scale datacenter renders defects more prevalent. When combined with the inevitable operational failures in mechanical devices, this increasingly results in unrecoverable data loss. Data protection in datacenters is now a critical issue. As far back as 2008, Google reported suffering a disk failure *every* time they executed a 6-hour long petabyte sort across 48,000 disks. In the case of correlated failure (i.e., due to the effect of physical conditions [9]), further aggravates the risk of data loss. A strategy to avoid it is to develop accurate prediction of impending disk failure events, so the data of the faulty disks can be copied to healthy disks with suitable lead time.

Since 1994, disk manufacturers developed self-monitoring technology to predict failures early enough to allow users to back-up their data [Hughes et al. 2002]. The Self-Monitoring and Reporting Technology (SMART) system uses attributes collected during normal operation (and during off-line tests) to set a failure prediction flag. The SMART flag is a signal read by operating systems and third-party software to warn users of impending drive failure. Some of the attributes it uses include counts of track seek retries, write faults, reallocated sectors, head fly heights, and high temperature. Most attributes are error count data, implying positive integer data values, and a pattern of increasing attribute values over time indicates impending failure. SMART essentially provides information about certain primitive disk health. Currently, all manufacturers use a threshold algorithm which triggers a SMART flag when any attribute exceeds a predefined value. The thresholds are set conservatively to avoid false alarms at the expense of predictive accuracy, with an acceptable false alarm rate on the order of 0.1%. For the SMART algorithm currently implemented in drives, manufacturers estimate the detection rate to be $3 - 10\%$. Since only thresholding the SMART statistics does not yield enough prediction accuracy, a large number of attempts have been made to improve the accuracy using various different machine learning models and other measures.

To improve SMART's failure prediction accuracy, [Hamerly and Elkan 2001] employed a Bayesian modeling approach. They tested on a dataset concerning 1936 drives, achieving failure detection rates of 55% for naive Bayes classifier at about 1% FAR. Another study was performed by [Hughes et al. 2002]. They used a Wilcoxon rank-sum test to build prediction models. They proposed two different strategies: multivariate test and ORing single attribute test. Their methods were tested on 3,744 drives. The highest detection rate achievable was 60% with 0.5% FAR. [Murray et al. 2004] compared the performance of SVM, unsupervised clustering, rank-sum test and reverse arrangements test. In subsequent work [Murray et al. 2005], they developed a new algorithm, termed multiple-instance naive Bayes (mi-NB). They found that, on the dataset concerning 369 drives, ranksum test outperformed SVM for certain small set of SMART attributes (28.1% failure detection at 0% false failure detection rate). When using all features, SVM achieved the best performance of 50.6% detection with 0% FAR. In [Zhu et al. 2013], the authors applied back-propagation learning on a feed-forward neural network to train on the SMART attributes, and achieved a failure detection rate of 95% with reasonably low FAR. In [Featherstun and Fulp 2010], the authors used syslog event records instead of the SMART attributes and used support vector machine to mine the syslog messages for disk failure prediction. They achieved approximate 80% accuracy. In [Pinheiro et al. 2007], a very large number of disks were considered in a Google datacenter and the authors presented a comprehensive analysis of the correlation between failures and parameters believed to affect disk lifetime. One fact they reported in [Pinheiro et al. 2007] is that the temperature bears no direct correlation with disk failure patterns.

In short, various predictive models have been built and reported in the literature to accurately predict the disk failure event with certain lead times. In datacenters, the implication of such predictions has not yet been systematically analyzed. In the event of a false prediction, the storage service provider incurs an extra cost and therefore the goal is to minimize the false prediction rate. On the other hand, to reduce the false prediction rate, the predictive models report low failure detection rate that in turn can result in a data loss event for large datacenter.

In this paper, we present as a baseline a systematic study of data loss in large datacenters when the disk failure prediction is not accurate. We also analyze the cost implication of false predictions. We then consider a generic form of predictive models behavior in terms of the receiver operating characteristic (ROC) curve, and show that for certain fixed cost, the effect on mean-time-to-data-loss (MTTDL) with respect to the predictive accuracy or the true failure detection rate. We find that the MTTDL follows a 'knee' of the curve with the prediction accuracy; i.e., after a certain predictive accuracy threshold, a small increase in accuracy yields a large gain in MTTDL. First we analyze the MTTDL for a datacenter using some generic protection mechanism. We then consider three protection case studies, namely RAID6, RAID triple parity, and Erasure Coding. We then analyze the implication of the true impending disk failure detection rate on the MTTDL. We also analyze the cost incurred due to false prediction. We analyze the implication of the impending disk failure prediction algorithms with empirical results. In doing the analysis of the impact of prediction accuracy of the disk failure algorithms for the datacenters, we make the following assumptions: (i) We consider that the datacenter as a large disk farm where the protection mechanism of all disks are same although in real datacenters, often this assumption may not be valid; (ii) We consider only the failures of disks – in practice, the disks are connected over network/nodes and the network/nodes themselves can fail which we do not include in our analysis; (iii) In this analysis, we assume the exponential failure distribution although in practice, the failure distribution deviates from the exponential distribution [Schroeder et al. 2016]. We consider the exponential distribution so that Markov model can be used to estimate the MTTDL for very large number of disks. For other distribution, no such theoretical model exists for large number of disks. As described in the paper [Elerath and Schindler 2014], certain more accurate models have been analyzed, however, it is difficult to extend that model for large number of disks. For large number of disks, only way to obtain MTTDL with failure distribution other than exponential, is to go through the simulation of the failure events. However, the simulation process does not scale for very large number of disks. We compare the theoretical model with simulation results on individual array; (iv) While considering the impact of wrong prediction, we analyze the disk replacement cost. The disk replacement cost is a quantitative number available within the company that includes the cost of maintenance center, people impact and the cost of the disk itself. In practice, there is an implication on SLA to the customer for data unavailability. Usually the SLA is agreed upon with the customer for disk replacement. In this analysis, we do not include the SLA conditions for predictive failure cases. In [Tai et al. 2014], the authors have analyzed a way of showing data replication mechanism without violating the SLA. In datacenter storage also, usually a certain percentage of the IO is reserved for the replication and rebuild operation so that customer workloads are not stopped completely during replication/rebuild operations.

Finally, the intent of this paper is to show how any impending disk failure prediction algorithm should be useful (or not at all useful) for a datacenter or more precisely to a large disk farm. We do not intend to compare the existing disk failure algorithms since (i) the prediction accuracies reported in the literature so far use different types of measurements and telemetry data (environment) to design the algorithms which may not be applicable uniformly to all algorithms, (ii) the amount of training data considered widely differ for different algorithms. So, it is very difficult to have an apple-to-apple comparison of the prediction algorithms only using the reported prediction accuracies. Therefore, the intent of this paper is not to experimentally compare the benefits of the existing algorithms in a large datacenter of a large disk farm, but to provide a theoretical guideline about what kind of prediction accuracy can be meaningful for datacenter to deploy the prediction technology.

2. RELIABILITY ANALYSIS FOR LARGE NUMBER OF DISKS

We first provide a generalized framework for analyzing the *mean time to data loss* (MTTDL) of a system of disks by generalizing the model of analysis as provided in [Paris et al. 2012, Paris et al. 2014]. Next we use the framework to estimate the MTTDL of different protection mechanisms. In general, disk failures may not follow the exponential distribution as studied in [Schroeder and Gibson 2007]. The operational failures can be better modeled by Weibull or Gamma distributions as studied in [Schroeder and Gibson 2007]. We assume that the disks fail independently. Once a disk fails, the recovery process immediately starts to recover the data. If multiple disks fail then the recovery of these disks can take place in parallel. We also assume that the recovery rate is exponentially distributed.

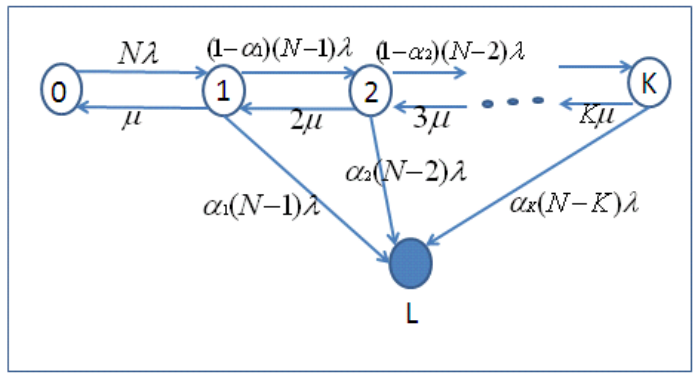

Figure 1: A continuous time Markov chain of the disk failures

Let $\lambda$ be the failure rate of the exponentially distributed failure process, $\mu$ be the recovery rate of the exponentially distributed recovery process. Figure 3 shows a continuous time Markov chain of the entire failure process for different failure rates. The system is initially in state 0 which indicates that there is no failure and all disks are operational. Since there are $N$ disks in the system, there can be $N\lambda$ probability of one disk failure. After one disk failure, the system goes to state 1. The failed disk can be replaced and data can be recovered with a recovery probability $\mu$. Once the data is recovered the system returns to state 0. In State 1, if another disk fails then there may be data loss with a probability $\alpha_1 \lambda$. Since there are (N-1) operational disks in State 1, there is a probability $(N-1)\alpha_1\lambda$ of data loss. If there is data loss, system moves to the data loss state $L$. If there is no data loss then system moves to State 2 with 2 disk failures with a probability $(1-\alpha_1)(N-1)\lambda$. Since both the failed disks can be replaced and recovered simultaneously, the system returns from State 2 to State 1 with a probability $2\mu$. Extending this logic, we can say that the system can move from a state $i$ to state $i+1$ with a probability $(1-\alpha_i)(N-i)\lambda$ and there is a probability $\alpha_i(N-i)\lambda$ of data loss. If the data loss happens then the system moves to data loss state $L$. The system can return from state $i+1$ to state $i$ with a probability $(i+1)\mu$. Let there be at most $K$ disk failures after which the system cannot be returned to normal state if one more disk fails, and data loss becomes inevitable. In that case, $\alpha_K = 1$ and the system goes to data loss state $L$ after one more disk fails. With constant rates of failure and recovery under the assumption of the exponentially distributed failure and recovery processes, we can express the transient changes in the probability of system states $P_0, P_1, P_2, P_3, \ldots, P_K$ as the Kolmogorov system of differential equations [Trivedi 2008] given by Equation (1).

$$\frac{dP_0(t)}{dt} = \mu P_1(t) - N\lambda P_0(t)$$

$$\frac{dP_1(t)}{dt} = N\lambda P_0(t) - ((N-1)\lambda + \mu)P_1(t) + 2\mu P_2(t)$$

$$\vdots$$

$$\frac{dP_{i+1}(t)}{dt} = (1-\alpha_i)(N-i)\lambda P_i - ((i+1)\mu + (N-(i+1))\lambda)P_{i+1}(t) + (i+2)\mu P_{i+2}(t)$$

$$\frac{dP_K(t)}{dt} = (1-\alpha_{K-1})(N-K+1)\lambda P_{K-1} - (K\mu + (N-K)\lambda)P_K(t)$$

$$\frac{dP_L(t)}{dt} = \sum_{i=1}^{K} \alpha_i(N-i)\lambda P_i \quad (1)$$

In Equation (2), $i+2 \leq K$, $P_0(0) = 1$, and $P_i(0) = 0$ for all $i > 0$. First we obtain the Laplace transformation of the system of equations. The mean time to data loss (MTTDL) can be expressed as [Trivedi 2008].

$$MTTDL = -\frac{dL}{ds}\bigg|_{s=0} \quad (2)$$

The loss in the system can be expressed as

$$L(s) = sP_L(s) = U(s)/V(s) \quad (3)$$

The loss function is derived from the continuous time Markov chain model in Laplace transformed space as described in [Trivedi 2008].

Where $U(s)$ and $V(s)$ are given as

$$V(s) = \beta_0(s) \quad (4)$$

Where

$$\beta_i = (s + i.\mu + (N-i)\lambda)\beta_{i+1} - (i+1)(N-i)(1-\alpha_i)\mu\lambda\beta_{i+2} \tag{5}$$

with $\beta_{K+1} = 1$, and $\beta_j = 0$ for all $j > K+1$.

The numerator $U(s)$ in Equation (3) is given as

$$U(s) = \sum_{i=1}^{K} \alpha_i \prod_{j=0}^{i}(N-j)\prod_{j=1}^{i-1}(1-\alpha_j)\lambda^{i+1}\beta_{i+1}(s) \tag{6}$$

Therefore, the loss function $L(s)$ is determined by $\beta(s)$ which in turn depends on the coefficients $\alpha$. Equating the negative of derivative of the loss function and letting $s=0$, we obtain the MTTDL for given values of $\alpha$.

Therefore,

$$MTTDL = \frac{V(0)\frac{dU(s)}{ds}\bigg|_{s=0} - U(0)\frac{dV(s)}{ds}\bigg|_{s=0}}{V(0)^2} \tag{7}$$

Since the derivatives are difficult to express in closed form equations, we programmatically compute the MTTDL using recursive formulations instead. Depending on the system configuration, the values of $\alpha$ change and the respective MTTDLs change.

### 2.1 Analysis for RAID6

In RAID6, an array of disks is protected by two parity disks. For example, in (8+2) RAID6, we have 8 data disks and two parity disks. We consider the general case of $l$ arrays of $(n+2)$ RAID6. RAID6 can protect against any two disk failures in an array. Therefore, we have $\alpha_1 = 0$. If there are three disks failure in a single array then there is data loss. Therefore, we have

$$\alpha_2 = \frac{\binom{n+2}{3}l}{\binom{N}{3}} \tag{8}$$

where $N = (n+2)l$ \hfill (9)

The numerator of Equation (8) is derived by considering the possible combination of data loss for three disk failures in one single array and then generalizing it to $l$ arrays. Since there are $n$ data disks in one array with 2 parity disks, it is the possible combinations for one array multiplied by $l$. The total number of disks is $N = (n+2)l$. Therefore, the total number of possibilities of three disk failures is given in the denominator which include the cases of data loss events and no data loss events. We extend the same logic for all cases subsequently.

In the case of four failures, we can select any one disk in addition to the configurations for three disk failure, and those results into data loss. Therefore,

$$\alpha_3 = \frac{(N-3)\binom{N}{3}\alpha_2}{\binom{N}{4}} = 4\alpha_2 \tag{10}$$

In general, for any subsequent failure, we have

$$\alpha_{i+1} = \min(1,(i+2)\alpha_i) \text{ for } i \geq 3. \tag{11}$$

## 2.2 Analysis for RAID Triple Parity

We consider that there are $l$ arrays of $(n + 3)$ disks, each array comprising of $n$ data disks and 3 parity disks. If there are three disks failures in any array then also RAID-TP is able to recover the data. Therefore, we have

$$\alpha_3 = \frac{\binom{n+3}{4} l}{\binom{N}{4}} \tag{12}$$

where $N = (n+3)l$. For any subsequent failures, we have

$$\alpha_{i+1} = \min(1, (i+2)\alpha_i) \text{ for } i \geq 4. \tag{13}$$

## 2.3 Erasure Code

We derive the MTTDL for $(n+m, n)$ Reed-Solomon code in the same way as RAID-DP and RAID-TP. An $(n+m, n)$ RS code can protect against $m$ disk failures but in the case of $m + 1$ disk failures there is a data loss.

We consider that there are $l$ arrays of $(n+m)$ disks, each array comprising of $n$ data disks and $m$ extra disks (similar to parity disks) having the RS codes. If there are $m$ disk failures then the RS code is able to recover the data. Therefore,

$$\alpha_1 = \alpha_2 = \ldots = \alpha_{m-1} = 0 \tag{14}$$

If there is $(m+1)$th disk failure then we have

$$\alpha_m = \frac{\binom{n+m}{m+1} l}{\binom{N}{m+1}} \tag{15}$$

where $N = (n+m)l$. For any subsequent disk failures, we have

$$\alpha_{i+1} = \min(1, (i+2)\alpha_i) \text{ for } i \geq m+1. \tag{16}$$

## 3. PREDICTION ANALYSIS OF DISK FAILURES

### 3.1 ROC Curve

In predicting the disk failure events using machine learning techniques, two parameters are used to evaluate the prediction algorithm namely, the true positive rate (TPR) and false positive rate (FPR), and are defined as follows:

$$TPR = \frac{number\ of\ disks\ correctly\ predicted\ to\ be\ failed}{Total\ number\ of\ disks\ actually\ failed} \tag{17}$$

$$FPR = \frac{number\ of\ disks\ incorrectly\ predicted\ to\ be\ failed}{Total\ number\ of\ disks\ that\ did\ not\ fail} \tag{18}$$

Evidently the most desired prediction performance is TPR =1 and FPR = 0. In practice, the prediction algorithms are not ideal. If every fault event needs to be predicted then there are some false predictions. Whenever a false prediction happens, the perfect disk is replaced. Every replacement of a disk incurs some cost. Therefore, if the false positive rate is very high then it implies a very high maintenance cost to the data center. On the other hand, if we maintain a very low FPR then the resulting TPR may also be low. A low TPR may result into significant data loss that is even more crucial for a datacenter.

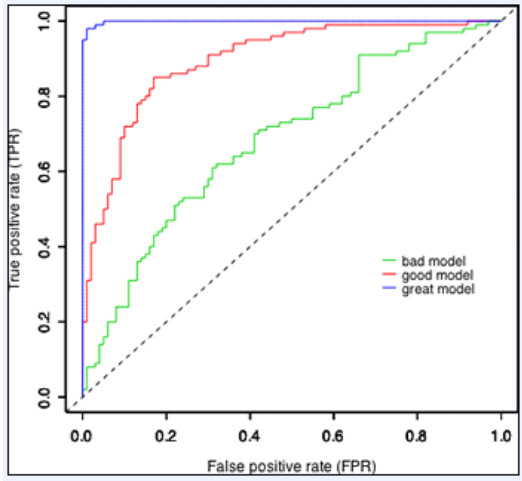

**Figure 2: A typical ROC curve**

In general, the FPR vs. TPR curves are known as "receiver operating characteristic" (ROC) curve [Weiss 2006]. A typical ROC curve [Weiss 2006] is shown in Figure 2. For any random prediction algorithm that predicts failures uniformly distributed among the positive and negative samples, true positive rate becomes equal to false positive rate. Therefore the diagonal line in the ROC represents the performance of a truly unbiased random algorithm. If the ROC is above the diagonal line then the algorithm is not truly random and carries some predicting capability. An ideal algorithm will have TPR = 1 for all FPR, $0 \leq FPR \leq 1$. For any realistic technique TPR increases as the FPR increases. Initially TPR increases very sharply with FPR and then the rate of increase in TPR slows down as the FPR increases.

The performance of a prediction algorithm is often characterized by the "Area Under the ROC Curve" (AUC). For any truly unbiased random algorithm, AUC = 0.5. On the other hand, for an ideal algorithm, AUC = 1. We empirically model the ROC curve as

$$TPR^p + (1-FPR)^p = 1 \qquad (19)$$

where $p \geq 1$ is a constant, and $0 \leq FPR \leq 1, 0 \leq TPR \leq 1$. If p = 1 then the ROC corresponds to a random unbiased algorithm with TPR=FPR. As $p \to \infty$, ROC corresponds to the ideal algorithm. We will analyze the implication of disk failure prediction algorithms for different AUC considering the empirical parametric form as in Equation (19).

### 3.2 Implication of TPR and FPR

For any disk failure correctly predicted, it has implication in the reduction of the data loss. Here we restrict ourselves to the data loss that happens due to disk failures. If there is a disk failure followed by an operational failure in the same RAID group during the reconstruction process then it implies to data loss event. If the failure event can be predicted successfully well ahead before the failure then the disk can be replaced with the necessary replication process of the stored data. In such a scenario, the data loss that can happen due to subsequent failures can be protected. In other words, for every correct prediction there is a reduction in the effective disk failure rate (since the disk is replaced in the operational condition with another new disk). In our analysis, we considered the failure rate as $\lambda$ with exponential distribution where the disks fail independently. If the failing disks are predicted to fail correctly then the effective failure rate becomes $\lambda(1-TPR)$. When TPR = 1 i.e., all failure events are correctly predicted then all failing disks are replaced in the operational condition and there is no data loss. On the other hand if TPR = 0 then no failing disks are replaced and the failure rate remains the same as the original. We can therefore, derive the effect of correct failure prediction on the MTTDL using the modified failure rate $\lambda(1-TPR)$.

Any incorrect failure prediction incurs an extra cost to the datacenter. In any datacenter, a disk that has not failed is replaced after a certain time period T (usually the average life-span of a disk). For example, in many enterprises, a disk is replaced after five years, i.e., T = 5 years. Let us assume that the datacenter has $N$ disks. Therefore, if the present time instant is t = 0 then all $N$ disks are operational from time t = -T to t= 0. Let us assume that the installation time of the disks (in new condition) is uniformly distributed over t = -T to t = 0. Therefore, the total number of disks failing in this period (assuming exponential distribution of failure) can be computed as the expected value that a disk will fail during a period t = 0 to t=T multiplied by the total number of disks since the installation period of disks are uniformly distributed. Therefore, the total number of disks that are failed over a period of T can be expressed as $N(1-e^{-\lambda T})$. Therefore, the total number of disks failing per unit time is

$$\frac{N}{T}(1-e^{-\lambda T}) \tag{20}$$

At any instant, the total number of disks not failing is given as

$$N - \frac{N}{T}(1-e^{-\lambda T}) \tag{21}$$

Therefore, the additional non-faulty disks that are predicted to be failed (and replaced) is given as

$$FPR\left[N - \frac{N}{T}\left(1-e^{-\lambda T}\right)\right] \tag{22}$$

Let $c$ be the cost of replacing one disk then the total additional cost is given as

$$Cost = c \times FPR\left[N - \frac{N}{T}\left(1-e^{-\lambda T}\right)\right] \tag{23}$$

For example, if FPR = 0 then there is no additional cost. On the other hand, if FPR = 1, then all non-failing disks are replaced. In other words, all disks are replaced at a particular unit time. In our analysis, we consider the unit time as the time to replace a failing disk including all logistics. For example, whenever a prediction about a possible failure happens, the admin needs to check the inventory for new disks. If it is not available in the local inventory then it has to be imported from another inventory. Then the data in the existing disk need to be copied on the new disk. Finally the existing disk needs to be replaced with the new disk. If the entire process takes 6 hours then the prediction must happen before 6 hours of actual failure. The disk replacement cost $c$ includes the cost of the new disk, the cost of copying the existing data onto the new disk, the labor cost of replacing the existing disk with the new disk, the cost of maintenance center of rechecking (and possible repair) the faulty disk. An approximate order of the cost is estimated as $c = \$375$ (available from the company's internal process). In practice, the disk replacement also involves SLA to the customer. Usually for data replication and rebuild a fraction of the IO is dedicated to this purpose apart from the usual workload IO. We do not consider the SLA in our present analysis since it is very difficult to quantify that factor.

4. **ANALYSIS RESULTS**

In this section, we analyze the results of implication of failure prediction. We first consider a data protection mechanism using RAID6. We consider (8+2) RAID6 configuration and there are 10000 such RAID groups resulting into 80,000 data disks and 20,000 parity disks in a datacenter totaling 100,000 disks. We consider the average life-span of a disk as 5 years such that the average rate of failure in the exponentially distributed failure process is $\lambda = 1/(1825 \times 24)$ per hour. We vary the mean time to recovery (MTTR) from 5 hours to 10 hours. Figure 3 shows the mean time to data loss (MTTDL) in terms of the number of days as we increase the TPR (the FPR can be assumed to be constant and does not affect the MTTDL). From Figure 3, we observe that initially MTTDL is very flat and does not increase much even if we increase the TPR. As the TPR becomes close 0.78, the MTTDL rises above five years for an MTTR = 5 hours. On the other hand, for an MTTR = 10 hours, TPR should be greater than 0.88 in order to have an MTTDL greater than 5 years. Note that in a large datacenter and MTTR = 5 hours is an aggressive estimate whereas an MTTR = 10 hours is more reasonable. So with reasonable maintenance conditions, a failure prediction algorithm should be able to predict with almost 90% true positive rate in order to maintain at least

5 years of mean time to data loss for a large datacenter having 100,000 disks using RAID6 protection mechanism. Figure 3 reveals that MTTDL for a large datacenter changes with the TPR in a similar way to the "hockey-stick" curve where latency changes with the IOPS in a storage system. This behavior is a general behavior irrespective of the failure distribution of the disks. This happens due to the fact that as TPR increases, the effective mean time to failure increases. When TPR becomes 100%, the mean time to failure becomes infinity (always the failing disks are replaced) and MTTDL becomes infinity or there will be no data loss. This very fact is true irrespective of the failure distribution of the disks. Depending on the number of disks, failure distribution and the protection mechanism, there will be a cut-over point on TPR for which MTTDL will be above a certain threshold and the MTTDL rises to infinity and the TPR goes to 100%.

As expressed in Equation (23), the additional cost to company due to wrong prediction is linearly proportional to the FPR. We show the dependency of the additional cost on FPR in Figure 4 where we assume that c = $375. The additional cost is simply linear with the false positive rate of prediction, the slope of which depends on the size of the datacenter (i.e., the number of disks present in the datacenter) and the cost of replacing each disk. Figure 4 is just a linear curve that represents the simple cost model that we consider.

We next analyze the implication of additional cost on MTTDL. We consider the parametric form of the ROC curve as in Equation (19). From Figures 5 and 6, we see that with the increase in AUC, there is a sharp increase in MTTDL for the same additional cost. Even for the very high values of AUC (such as above 0.95), a small increase in AUC results in significant increase in the MTTDL.

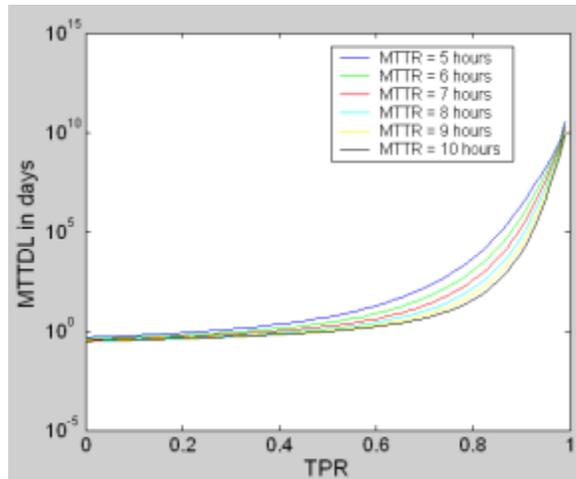

**Figure 3: MTTDL vs. True positive rate of failure prediction**

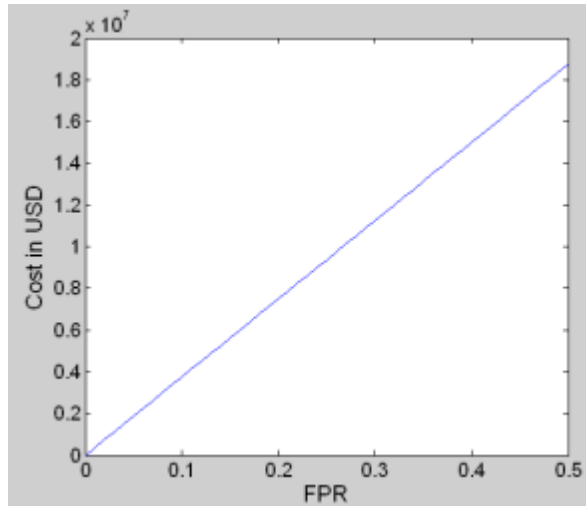

**Figure 4: Additional cost vs. false positive rate of failure prediction**

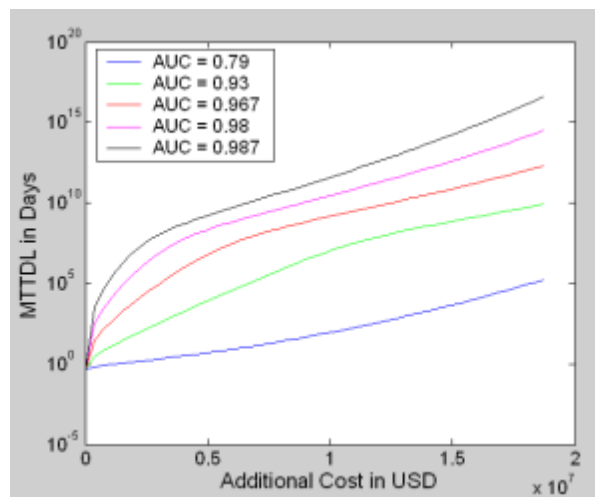

**Figure 5: MTTDL vs. Additional cost for different AUC of failure prediction**

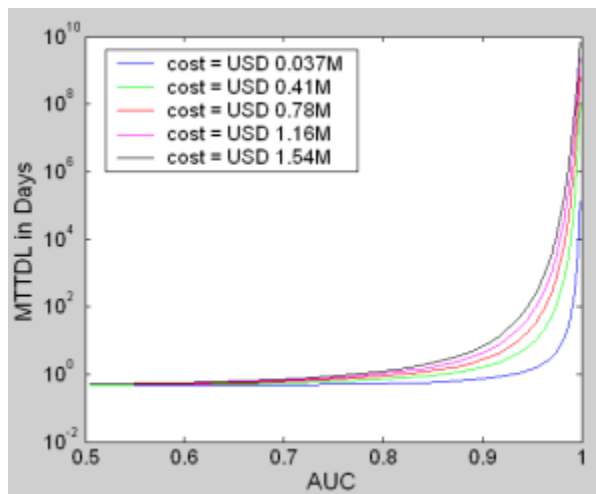

**Figure 6: MTTDL vs. AUC of prediction for different additional costs**

In Table 1, we show the effect of TPR of disk failure prediction algorithms on the MTTDL for various data protection mechanisms including RAID6, RAID triple parity and (20,16) RS code. In all cases, we considered 80000 data disks. We keep the number of data disks to be same in all three cases (i.e., 80,000 data disks) and the number of parity disks changes depending on the protection mechanism. Our intent to keep the capacity same to ensure the datacenter is capable of storing the same amount of data in all three cases. For (8+2) RAID6, the datacenter requires 20,000 parity disks; for (8+3) RAID-TP, the datacenter requires 30,000 parity disks additionally, and for (20,16) erasure code, the datacenter requires 20,000 parity disks which is the same as in the case of RAID6.

Next we present certain simulation results on the MTTDL. We used the disk failure events in reported in the company install-base and the data has been analyzed in [Magie et al. 2015]. The analysis revealed that enterprise near-line class drives have approximately constant failure rates governed by exponential failure distribution. Relatively low-cost archival hard disk drives are subject to early-life decreasing failure rate and a subtle and significant wear-out problems governed by bathtub curves that is a resultant of the mixture of exponential and Weibull distribution. Within the company, there are extensive simulation of the failures of disk drives validated from the field data (most of which cannot be revealed due to confidentiality agreement with the customers). The best fit of the distribution is a combination of exponential and Weibull distribution, representing a drive that, in addition to its intrinsic failure mechanisms, also has a wear out failure mechanism [Schroeder and Gibson 2007]. An exponential distribution with a mean-time-to-failure of 5 years and a Weibull distribution with a scale parameter of 70,000 hours and a shape parameter of 4.0 were used in the simulations. Considering the data protection mechanisms deployed in the various customer bases, the simulation results are available for (16+2) RAID-DP [Corbett et al. 2004] and (24+3) RAID-Triple Parity [Goel and Corbett 2012]. We compare the MTTDL of single array of (16+2) and (24+3) configurations with our theoretical models. In the simulation, MTTDL of (16+2) RAID-DP is reported to be $2.083*10^6$ days whereas in the theoretical model we obtain $2.519*10^6$ days considering mean-time-to-failure as 5 years and MTTR as 24 hours. With the same simulation model and drive parameters, the simulation results reported within the company for (24+3) RAID-TP was $4.63*10^7$ days whereas the theoretical model yielded $1.61*10^8$ days which is roughly 3.5 times more than the simulation results available. Note that, the simulation (in the company) was performed using a mixture of exponential and Weibull distribution to fit the failure characteristics of the second class of drives.

5. **SUMMARY AND CONCLUSIONS**

In this paper, we presented a systematic analysis of the significance of the disk failure prediction algorithms. We analyzed the effect of true positive rate in prediction on the MTTDL in a large data center. We also analyzed the additional cost penalty against the false positive rate of prediction. An interesting outcome of this analysis is that the effectiveness of the prediction algorithms suddenly increases in terms of MTTDL beyond a certain true positive rate. In other words, the TPR vs. MTTDL curve has a 'knee' that is very similar to the latency vs. IOPS curve. This nature of the MTTDL vs. TPR curve is for all three data protection mechanisms that we analyzed. This analysis also shows what kinds of TPR and FPR are expected for a data center to maintain a certain threshold of MTTDL against a certain amount of budget leading to an expectation setting of the designed disk failure prediction algorithms.

**Table 1: MTTDL vs. MTTR for different TPR with different data protection mechanisms**

| TPR | MTTR (hours) | MTTDL (in days) | | |
|---|---|---|---|---|
| | | RAID6 (8+2) | RAID-TP (8+3) | RS Code (20,16) |
| 80 | 5 | 192.94 | $7.5 \times 10^4$ | $5.85 \times 10^7$ |
| | 10 | 1.56 | 39.73 | $3.51 \times 10^3$ |
| | 15 | 0.27 | 1.51 | 31.32 |
| 85 | 5 | $2.61 \times 10^3$ | $3.19 \times 10^6$ | $5.74 \times 10^9$ |
| | 10 | 12.5 | 986 | $2.22 \times 10^5$ |
| | 15 | 1.13 | 18.14 | 1073 |
| 90 | 5 | $9.31 \times 10^4$ | $5.96 \times 10^8$ | $2.67 \times 10^{12}$ |
| | 10 | 385 | $1.5 \times 10^5$ | $1.17 \times 10^8$ |
| | 15 | 18.75 | $1.48 \times 10^3$ | $3.33 \times 10^5$ |
| 95 | 5 | $6.91 \times 10^6$ | $3.06 \times 10^{11}$ | -- |
| | 10 | $1.86 \times 10^5$ | $1.19 \times 10^9$ | $5.94 \times 10^{12}$ |
| | 15 | $7.82 \times 10^3$ | $9.6 \times 10^6$ | $1.72 \times 10^{10}$ |